\documentclass{ws-ijmpcs}

\usepackage{amsmath}
\usepackage{amssymb}
\usepackage{bbm}
\usepackage{multirow}
\usepackage{setspace}
\usepackage{graphicx}
\usepackage{epsfig}
\usepackage{dcolumn}
\usepackage{bm}
\usepackage{subfigure}
\usepackage{slashed}
\usepackage{longtable}
\usepackage[colorlinks,
            linkbordercolor={0 0 0},
            pdfborder={0 0 0},
            linkcolor=blue,
            citecolor=blue,
            urlcolor=blue,
            breaklinks=true]{hyperref}

\def\rmsmall #1{\mbox{\scriptsize #1}}

\def\eqarray#1{\begin{eqnarray} #1 \end{eqnarray}}

\newcommand{\bea}{\begin{eqnarray}}
\newcommand{\eea}{\end{eqnarray}}
\newcommand{\be}{\begin{equation}}
\newcommand{\ee}{\end{equation}}

\newcommand{\la}{\label}

\newcommand{\ba}{\begin{eqnarray}}
\newcommand{\ea}{\end{eqnarray}}

\begin{document}

\markboth{K.F. Liu}
{Quarks and glue components of the proton spin}

\catchline{}{}{}{}{}
%
%%%%%%%%%%%%%%%%%%%%%%%%%%%%%%%%%%%%%%%%%%%%%%%%%%%%%%%%%%%%%%%%%%%%

\title{Quark and Glue Components of the Proton Spin from Lattice Calculation\footnote{Plenary talk presented
at SPIN2014, the 21st International Symposium on Spin Physics.}}

\author{Keh-Fei Liu\footnote{On behalf of $\chi$QCD Collaboration}}

\address{Department of Physics and Astronomy, University of Kentucky\\
Lexington, KY 40506 USA\footnote{liu@pa.uky.edu}\\
}

%\author{Second Author}

%\address{Group, Laboratory, Address\\
%City, State ZIP/Zone, Country\\
%second\_author@domain\_name}

\maketitle

%\begin{history}
%\received{Day Month Year}
%\revised{Day Month Year}
%\published{Day Month Year}
%\end{history}

\begin{abstract}
The status of lattice calculations of the quark spin, the quark orbital angular
momentum, the glue angular momentum and glue spin in the nucleon is summarized. The quark spin calculation
is recently carried out from the anomalous Ward identity with chiral fermions and is found
to be small mainly due to the large negative anomaly term which is believed to be the source of the 
`proton spin crisis'. We also present the first calculation of the glue spin at finite nucleon momenta.
\keywords{quark spin, glue spin, proton spin}
\end{abstract}

\ccode{PACS numbers: 12.38.Gc,11.40.Ha,12.38.-t,14.20.Dh}

\section{Introduction}	   \label{introduction}

\hspace{0.5cm}
Apportioning the spin of the nucleon among its constituents of quarks and glue is one of the most challenging issues in QCD both experimentally and theoretically. 

%\medskip

Since the contribution from the quark spin is found to be small
($\sim$25\% of the total proton spin) from the global analysis of deep inelastic scattering 
data~\cite{deFlorian:2009vb}, it is expected that the remainder should come from glue spin and the orbital angular momenta of quarks and glue. The quark spin contribution from $u$,~$d$~and~$s$ has been studied on the 
lattice~\cite{Dong:1995rx,Fukugita:1994fh} since 1995 with quenched approximation or 
with heavy dynamical fermions~\cite{Gusken:1999as}.\ Recently,\ it has been 
carried out with light dynamical 
fermions~\cite{QCDSF:2011aa,Engelhardt:2012gd,Abdel-Rehim:2013wlz,Babich:2010at}
 for the strange quark.\ We will report the calculation of both the connected insertion (CI) and disconnected 
 insertion (DI) contributions to quark spin from $u$,\ $d$,\ $s$ and 
$c$ using anomalous Ward identity from the overlap fermion~\cite{Yang:2015xga}.

%\medskip

As for the quark orbital angular momenta,\ lattice calculations have been carried out for the
connected insertions 
(CI)~\cite{Mathur:1999uf,Hagler:2003jd,Gockeler:2003jfa,Brommel:2007sb,Bratt:2010jn,Alexandrou:2011nr,Syritsyn:2011vk,Alexandrou:2013joa}.\ They are obtained by subtracting 
the quark spin contributions from those of the quark angular momenta.\ It has been shown 
that the contributions from $u$ and $d$ quarks almost cancel each other.\ Thus for connected 
insertion,\ quark orbital angular momenta turn out to be small in the quenched 
calculation~\cite{Mathur:1999uf,Gockeler:2003jfa} and nearly zero in dynamical fermion 
calculations~\cite{Brommel:2007sb,Bratt:2010jn,Alexandrou:2011nr,Syritsyn:2011vk,Alexandrou:2013joa}.
\ On the other hand,\ gluon helicity distribution $\Delta G(x)/G(x)$ from 
COMPASS,\ STAR,\ HERMES and PHENIX experiments is found to be close to 
zero~\cite{Adolph:2012ca,Djawotho:2011zz,Airapetian:2010ac,Stolarski:2010zz,Adare:2008qb}.\ 
A global fit~\cite{Aschenauer:2013woa} with the inclusion of the polarized deep 
inelastic scattering (DIS) data from COMPASS~\cite{Alekseev:2010hc} and the 2009 data from 
RHIC~\cite{Aschenauer:2013woa},\ gives a glue contribution 
$\displaystyle\int_{0.05}^{0.2} \Delta g(x) dx = 0.1 \pm_{0.07}^{0.06}$ to the total proton spin 
of $1/2 \hbar$ with a sizable uncertainty.\ Most recent analysis~\cite{deFlorian:2014yva} of high-statistics 
2009 STAR~\cite{Adamczyk:2014ozi} and PHENIX~\cite{Adare:2014hsq} data show an evidence of non-zero glue helicity in the proton. For $Q^2=10$ ${\rm GeV}^2$, they found the gluon helicity distribution $\Delta g(x,Q^2)$ positive and away from zero in the momentum fraction range $0.05\leq x \leq0.2$. 
However, the result presented in~\cite{florian} has very large uncertainty in the small $x$-region. Moreover,\ it is argued based on analysis of single-spin asymmetry in unpolarized lepton scattering from a transversely polarized nucleon that the glue orbital angular momentum is absent~\cite{Brodsky:2006ha}.\ Given that
DIS experiments and quenched lattice calculation thus far reveal that only $\sim 25\%$ of the proton spin 
comes from the quark spin,\ lattice calculations of the orbital angular momenta show that the connected insertion (CI) parts have negligible contributions,\ and gluon helicity from the latest global analysis~\cite{deFlorian:2014yva} is $\sim 40\%$ albeit with large error, there are still missing components in the proton spin.
In this context, it is dubbed a `Dark Spin' conundrum~\cite{Liu:2012nz,Deka:2013zha}.

    In this talk, I shall present a complete decomposition of the nucleon spin in terms of the quark spin,
the quark orbital angular momentum, and the glue angular momentum in a quenched lattice calculation.
I will then summarize the lattice effort in calculating the strange quark spin in dynamical fermions and
present a result of the total quark spin from a lattice calculation employing the anomalous Ward identity
and, finally, I will show a preliminary first calculation of the glue spin at finite nucleon momenta.

\section{Formalism}   \label{formalism}

\hspace{0.5cm}
         It is shown by X. Ji~\cite{Ji:1996ek} that there is a gauge-invariant separation of the proton spin operator into
the quark spin, quark orbital angular momentum, and glue angular momentum operators
\begin{equation}       
\vec J_{\rmsmall{QCD}} = \vec J_q + \vec J_g
  = \frac{1}{2} \vec\Sigma_q + \vec{L}_q + \vec{J}_g ,
\label{ang_op_def_split_2}
\end{equation}
where the quark and glue angular momentum operators are defined from the
symmetric energy-momentum tensor
\begin{equation}
  J_{q,g}^i 
 = \frac{1}{2}\,\epsilon^{ijk}\,\int \, d^3x\, \left(\mathcal{T}_{q,g}^{0k}\, x^j
    - \mathcal{T}_{q,g}^{0j}\, x^k\right) ,
\label{ang_op_def_split_1}
\end{equation}
with the explicit expression
\begin{equation}
  \vec{J}_q = \frac{1}{2} \vec\Sigma_q + \vec{L}_q  =  \int d^3x \, 
  \bigg{[} \frac{1}{2}\, \overline\psi\,\vec{\gamma}\,\gamma^5 \,\psi 
 + \psi^\dag \,\{ \vec{x} \times (i \vec{D}) \} \,\psi \bigg{]} ,
\label{quark_ang_op_split_1}
\end{equation} 
for the quark angular momentum which is the sum of quark spin and orbital angular
momentum, and each of which is gauge invariant. The glue angular momentum
\begin{equation}
\vec{J}_g = \int d^3x \,\bigg{[} \vec{x} \times ( \vec{E} \times \vec{B} )\bigg{]} ,
\label{gluon_ang_op_def_split_1}
\end{equation}
is also gauge invariant. However, since it is derived from the symmetric energy-momentum
tensor in the Belinfante form, it cannot be further divided into the glue spin and orbital
angular momentum gauge invariantly. 

    Since the quark orbital angular momentum and glue angular momentum operators in
Eqs.~(\ref{quark_ang_op_split_1}) and (\ref{gluon_ang_op_def_split_1}) depends on
the radial vector $\vec{r}$, a straight-forward application of the lattice calculation is complicated 
by the periodic condition of the lattice,\ and may lead to wrong results~\cite{Wilcox:2002zt}.\ 
Hence,\ instead of calculating $J_q$ and $J_g$ directly,\ we 
shall calculate them from the energy-momentum form factors in the nucleon.

   The Euclidean energy-momentum operators for the quark and glue are
\begin{eqnarray}
\label{eq:q_contrib_def_2}
  {\mathcal T}_{\{4i\}q}^{(E)} 
  &=&  (-1)\, \frac{i}{4}\displaystyle\sum_f \overline {\psi}_f 
  \left[ 
    \gamma_4 \stackrel{\rightarrow} D_i 
    + \gamma_i \stackrel{\rightarrow} D_4
    -  \gamma_4 \stackrel{\leftarrow} D_i
    - \gamma_i \stackrel{\leftarrow} D_4
    \right] \psi_f , \\
\label{eq:g_contrib_def_2}
  {\mathcal T}_{\{4i\}g}^{(E)} 
  &=&  (+i)\, \bigg[-\frac{1}{2} \displaystyle\sum_{k=1}^3 2\, 
    \mbox{Tr}^{\rmsmall{color}} \left[G_{4k}\, G_{ki} + G_{ik}\, G_{k4} \right]\bigg] .
\end{eqnarray}
 where we use the Pauli-Sakurai representation for the gamma matrices and the covariant
 derivative is the point-split lattice operator involving the gauge link $U_{\mu}$. For the
 gauge field tensor $G_{\mu\nu}$, we use the overlap fermion Dirac operator. The connection
 between  $G_{\mu\nu}$ and the overlap Dirac operator has been derived~\cite{Liu:2007hq,Alexandru:2008fu}
\begin{equation}
  \mbox{Tr}_s \left[\sigma_{\mu\nu} D_{\rmsmall{ov}}(x,x)\right] 
  =  c_T\, a^2\, G_{\mu\nu}(x) + {\mathcal O}(a^3) ,
\label{Glue_Tensor}
\end{equation}
where $\mbox{Tr}_s$ is the trace over spin.\ $c_T = 0.11157$ is the proportional constant 
at the continuum limit for the parameter $\kappa = 0.19$ in the Wilson kernel of the overlap 
operator which is used in this work. The overlap Dirac operator $D_{ov}(x,y)$ is exponentially
local and the gauge field $G_{\mu\nu}$ as defined in Eq.~(\ref{Glue_Tensor}) is chirally smoothed
so that it admits good signals for the glue momentum and angular momentum in the lattice 
calculation~\cite{Deka:2013zha}.
     
The form factors for the quark and glue energy-momentum tensor are defined as 
\begin{eqnarray}
&&\langle p', s'  | {\mathcal T}_{\{4i\}q,g}^{(E)} | p, s\rangle
  = \left(\frac{1}{2}\right) \bar{u}^{(E)}(p',s') \left[T_1(-q^2)(\gamma_4\bar{p}_i 
   +  \gamma_i\bar{p}_4)\right. \nonumber\\
&&\,\,\,\,\,\,- \frac{1}{2m}T_2(-q^2)(\bar{p}_4 \sigma_{i\alpha} q_{\alpha} 
   +  \bar{p}_i \sigma_{4\alpha} q_{\alpha})
  - \left.\frac{i}{m} T_3(-q^2) q_4 q_i\right]_{q,g}\!\!\!\!\!\! u^{(E)}(p,s) .
\label{mat_element_2}
\end{eqnarray}
where the normalization conditions for the nucleon spinors are
\begin{equation}
\bar{u}^{(E)}(p,s)\, u^{(E)}(p,s)\, =\, 1\, , \, 
\hspace*{0.5cm}
\displaystyle\sum_s  u^{(E)}(p,s)\, \bar{u}^{(E)}(p,s)\, 
= \, \frac{\slashed{p} + m}{2m} , 
\label{norm_cond_euc}
\end{equation}

\subsection{Sum rules and renormalization}   \label{sumrules}

\hspace{0.5cm}
The momentum and angular momentum fractions of the quark and glue depend 
on the renormalization scale and scheme individually, but their sums do not
because the total momentum and angular momentum of the nucleon are conserved. We shall use the
sum rules as the renormalization conditions on the lattice.

Substituting the energy-momentum tensor matrix elements in Eq.~(\ref{mat_element_2})
to the matrix elements which define the angular momentum in Eq.~(\ref{ang_op_def_split_1}) 
and a similar equation for the momentum, it is shown~\cite{Ji:1996ek} that

\begin{eqnarray}
  \label{ang_op_def_split_3}
  J_{q,g} &=& \frac{1}{2} Z_{q,g}^L \left[T_1(0) + T_2(0)\right]_{q,g} ,\\
  \langle x\rangle_{q,g} &=& Z_{q,g}^L T_1(0)_{q,g} .
\label{momentum_fraction}
\end{eqnarray}
where $Z_{q,g}^L$ is the renormalization constant for the lattice quark/glue operator.
$\langle x\rangle_{q,g}$ is the second moment of the unpolarized
parton distribution function which is the momentum 
fraction carried by the quark or glue inside a nucleon.\ The other form factor,\ 
$T_2(0)_{q,g}$,\ can be interpreted as the anomalous gravitomagnetic moment 
in analogy to the anomalous magnetic moment,\ 
$F_2(0)$~\cite{Teryaev:1999su,Brodsky:2000ii}.

\smallskip

\noindent
Since momentum is always conserved and the nucleon has a total spin of 
$\displaystyle\frac{1}{2}$,\ we write the momentum and angular momentum sum rules 
using 
Eqs.~(\ref{ang_op_def_split_2}),~(\ref{ang_op_def_split_3})~and~(\ref{momentum_fraction}),\ 
as
\eqarray{
  \label{eq:mom_sum_rule}
  \langle x\rangle_{q} +   \langle x\rangle_{g}\, =\, Z_q^L T_1 (0)_q + Z_g^L T_1 (0)_g  &=& 1 , \\
  J_q+ J_g\, =\, \frac{1}{2}\,  \bigg\{
  Z_q^L \left[ T_1 (0) + T_2 (0) \right]_q + Z_g^L \left[ T_1 (0)  + T_2 (0) \right]_g 
  \bigg\} &=& \frac{1}{2} . 
  \label{eq:ang_mom_sum_rule}
}
It is interesting to note that from Eqs.~(\ref{eq:mom_sum_rule}) and 
(\ref{eq:ang_mom_sum_rule}),\ one obtains that the sum of the $T_2(0)$'s for the 
quarks and glue is zero,\ i.e.
\eqarray{
Z_q^L T_2 (0)_q + Z_g ^L T_2 (0)_g &=& 0 .
\label{eq:T_2_sum}
}
We used these sum rules and the raw lattice results to obtain the lattice renormalization
constants $Z_g^L$ and $Z_g^L$ and then use perturbation~\cite{Glatzmaier:2014sya} to calculate the quark-glue
mixing and renormalization in order to match to the $\overline{MS}$ scheme at 2 GeV which preserves the sum rules. 

\subsection{Results of a lattice calculation with quenched approximation}

\begin{figure}[h]
\centering
\subfigure[]
{\rotatebox{0}{\includegraphics[width=0.44\hsize]{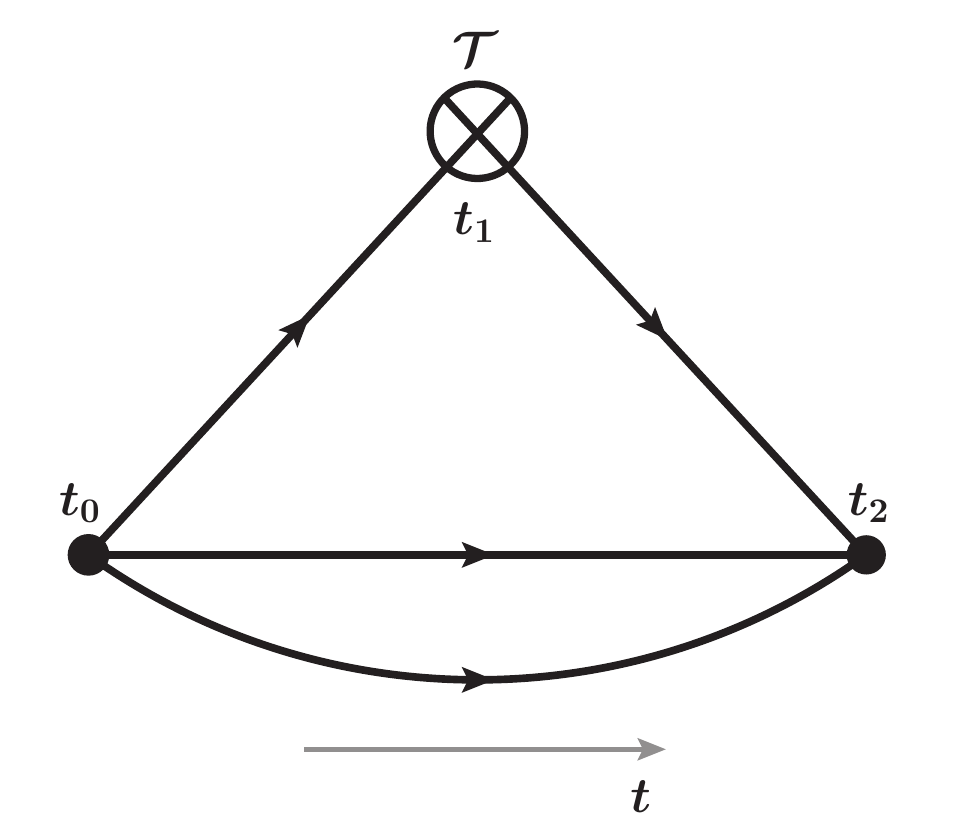}}
\label{fig:connected_insertion}}
\hspace{2mm}
\subfigure[]
{\rotatebox{0}{\includegraphics[width=0.44\hsize]{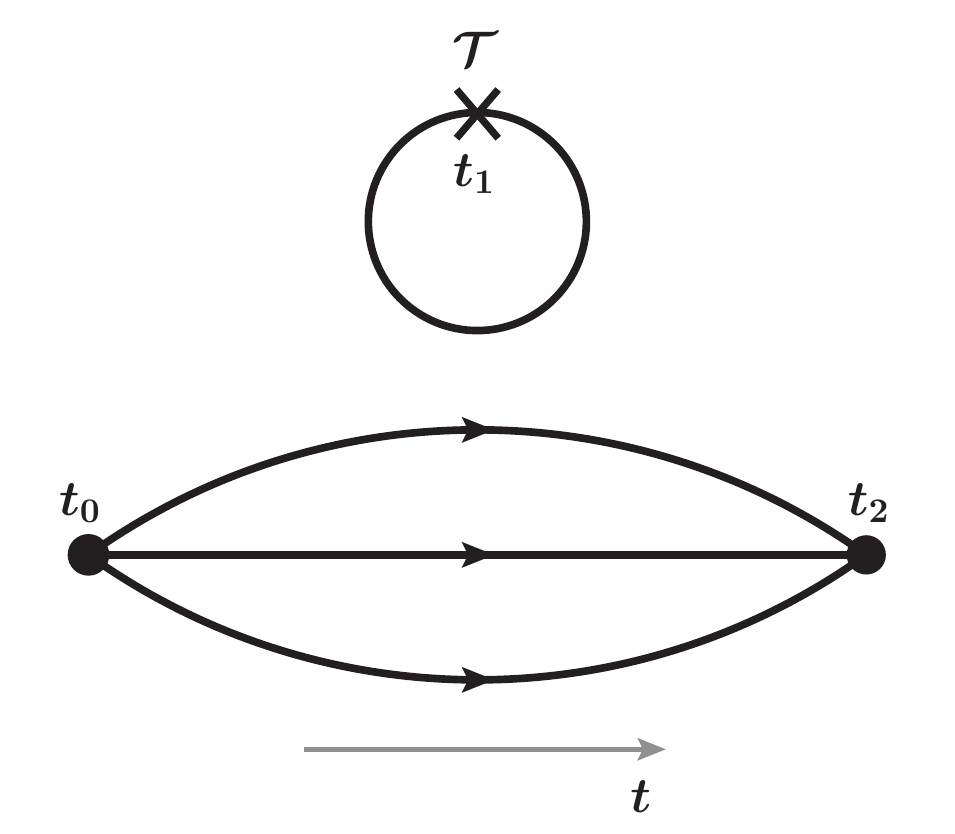}}
 \label{fig:disconnected_insertion}}
\caption{Quark line diagrams of the three-point function with current insertion in the Euclidean
         path integral formalism.\ 
         (a) Connected insertions (CI), and
         (b) disconnected insertions (DI).}
\label{fig:ci_and_di}
\end{figure}

\hspace{0.5cm}
Before we present the lattice results, we should point out that
the three-point functions for quarks which are needed to extract the form factors
in Eq.~(\ref{mat_element_2}) have two topologically distinct contributions in the
path-integral diagrams:\ one from connected insertions (CI) and the other from 
disconnected insertions (DI)~\cite{Liu:1993cv,Liu:1998um,Liu:1999ak,Liu:2012ch} (See 
Figs.~\ref{fig:ci_and_di}).\ They arise in different Wick contractions,\ and 
it needs to be stressed that they are not Feynman diagrams in perturbation theory.\ In 
the case of CI,\ quark/anti-quark fields from the operator are contracted with the 
quark/anti-quark fields of the proton interpolating fields.\ It represent the valence and
the higher Fock space contributions from the Z-graphs. In the case of DI,\ the 
quark/anti-quark fields from the operator contract themselves to form a current loop,\ which
represents the vacuum polarization of the disconnected sea quarks.

It should be pointed out that, although the quarks lines in the loop and the nucleon propagator appear
to be `disconnected' in Fig~\ref{fig:ci_and_di}(b),\  they are in fact correlated
through the gauge background fluctuation.\ In practice,\ the uncorrelated part of
the loop and the proton propagator is subtracted.\ The disconnected insertion (DI)
refers to the fact that the quark lines are disconnected. 
For the nucleon, the up and down quarks contribute to both CI and DI, while the 
strange and charm quarks contribute to the DI only.

      A quenched lattice calculation on has been carried out with 3 valence quark masses and
 extrapolated to the physical pion mass where the numerical details
 of the calculation are given~\cite{Deka:2013zha}. We shall present the results in the following table.

\begin{table}[htbp]
  \centering
  \renewcommand{\arraystretch}{1.4}
  \tbl{Renormalized results in $\overline{MS}$ scheme at $\mu = 2$~GeV.}
  { \begin{tabular}{|c||cc|cccc|}
    \hline
    & {\bf CI(u)} & {\bf CI(d)}  & {\bf CI(u+d)} &  {\bf DI(u/d)} & {\bf DI(s)} & {\bf Glue} \\
    \hline
    {\boldmath $\langle x \rangle$}
    & 0.413(38)  &  0.150(19) & 0.565(43) & 0.038(7) & 0.024(6) & 0.334(55) \\
    \hline
    {\boldmath $T_2(0)$} 
    & 0.286(108)  & -0.220(77) & 0.062(21) & -0.002(2) & -0.001(3) & -0.056(51) \\
    \hline
    {\boldmath $2J$} 
    &  0.700(123)  & -0.069(79) & 0.628(49) & 0.036(7) & 0.023(7) & 0.278(75)\\
    \hline
    {\boldmath $g_A$}
    &  0.91(11)  & -0.30(12)   & 0.62(9)  &  -0.12(1)  &  -0.12(1) & \--- \\
    \hline
    {\boldmath $2L$}
    &  -0.21(16)    &  0.23(15)   &  0.01(10)  &  0.16(1)  &  0.14(1) & \--- \\
    \hline
  \end{tabular}   \label{tab:chiral}
  }
\end{table}

For the unrenormalized lattice results,  
we find that $\left[T_2^u(0) + T_2^d(0)\right]$ (CI) is positive and $T_{2}^g (0)$ negative,\ 
so that the total sum including the small $\left[T_2^u (0) + T_2^d (0) + T_2^s (0)\right]$ (DI) 
can be naturally constrained to be zero (See Eq.~(\ref{eq:T_2_sum})) with the lattice normalization 
constants $Z_q^L = 1.05$ and $Z_g^L = 1.05$ close to unity.\ As discussed in Sec.~\ref{sumrules},\ the vanishing of  the total $T_2(0)$ is the consequence of momentum and angular momentum conservation. 
\begin{figure}[h]
  \centering
   \subfigure[] 
  {\rotatebox{0}%
    {\includegraphics[width=0.99\textwidth]{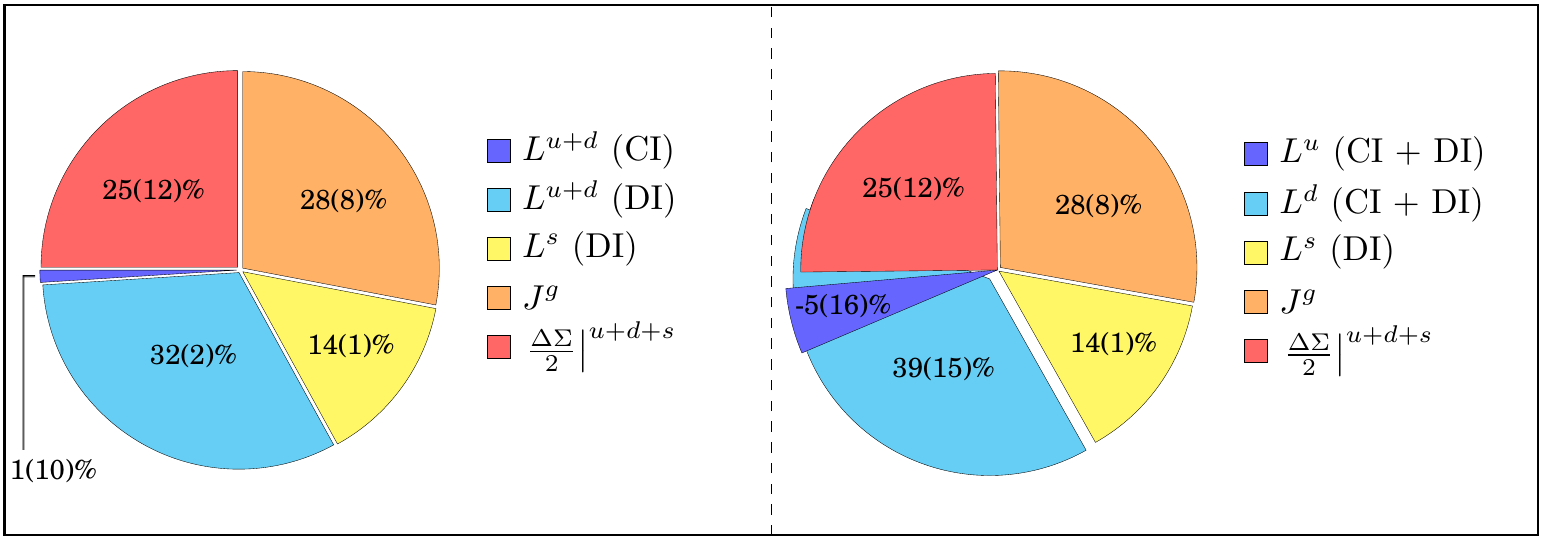}}
    \label{fig:pie_diag_orb_am}
  }
  \caption{Pie charts for the quark spin, quark orbital angular momentum and
  gluon angular momentum contributions to the proton spin.
        The left panel show the quark contributions separately for CI and DI,\ and the right panel 
    shows the quark contributions for each flavor with CI and DI summed together for $u$ and 
    $d$ quarks.%
  }
  \label{fig:pie_diag}
\end{figure}

The flavor-singlet $g_A^0$ which is the quark spin contribution to the nucleon has been 
calculated before on the same lattice~\cite{Dong:1995rx}.\ We can subtract it from the 
total quark angular momentum fraction $2J$ to obtain the orbital angular momentum fraction 
$2L$ for the quarks.\ As we see in Table~\ref{tab:chiral},\ the orbital angular momentum 
fractions $2L$ for the $u$ and $d$ quarks in the CI have different signs and they add up to 
zero,\ i.e.\ $0.01(10)$.\ This is the same pattern which has been seen with dynamical fermion configurations 
with light quarks which was pointed out in Sec.~\ref{introduction}.\ The large $2L$ for the $u/d$ 
and $s$ quarks in the DI is due to the 
fact that $g_A^0$ in the DI is large and negative,\ i.e.\ $-0.12(1)$ for each of the three 
flavors.\ All together,\ the quark orbital angular momentum constitutes a fraction of 
$0.47(13)$ of the nucleon spin.\ The majority of it comes from the DI.\ The quark spin fraction 
of the nucleon spin is $0.25(12)$ and the glue angular momentum contributes a fraction of 
$0.28(8)$.\ We show the quark spin, the quark orbital angular momentum and the glue
angular momentum in the pie chart in Fig.~\ref{fig:pie_diag}.\ The left panel shows the combination 
of $u$ and $d$ contributions to the orbital angular momentum from the CI and DI separately while 
the right panel shows the combined (CI and DI) contributions to the orbital angular momentum from 
the $u$ and $d$ quarks.

    Since this calculation is based on a quenched approximation which is known to contain
large uncontrolled systematic errors, it is essential to repeat this calculation with dynamical
fermions of light quarks and large physical volume.

\section{Quark spin from anomalous Ward identity}

    Attempts have been made to tackle the proton spin decomposition with light
dynamical fermions configurations. 
   There have been a number of calculations of the strange quark 
spin~\cite{QCDSF:2011aa,Engelhardt:2012gd,Abdel-Rehim:2013wlz,Babich:2010at}
which found the strange quark spin $\Delta s$ to be in the range from $-0.02$ to $-0.03$
which is several times smaller than that from a global fit of DIS and semi-inclusive DIS (SIDIS)
which gives $\Delta s \approx - 0.11$~\cite{deFlorian:2009vb}. The large negative contribution from
the strange quark is confirmed by a recent analysis~\cite{Leader:2014uua} 
of the world data on inclusive deep inelastic scattering data including COMPASS 2010 proton data on 
the spin asymmetries and the precise JLab CLAS data on the proton and deuteron spin structure functions
which gives $\Delta s + \Delta \bar{s} = - 0.106 \pm 0.023$~\cite{leader15}.

   Such a discrepancy between the global fit of experiments and the lattice calculation
of the quark spin from the axial-vector current has raised a concern that the renormalization constant 
for the flavor-singlet axial-vector current could be substantially different from that of the 
flavor-octet~\cite{Karsten:1980wd,Lagae:1994bv} at the lattice cutoff of $\sim 2$ GeV. 
The latter is commonly used for
the lattice calculations of the flavor-singlet axial-vector current for the quark spin. 
To alleviate this concern, we use the anomalous Ward identity (AWI) to calculate 
the quark spin~\cite{Yang:2015xga}. The anomalous Ward identity includes a triangle anomaly in the
divergence of the flavor-singlet axial-vector current
\begin{equation}\label{awi}
\partial^{\mu} A_{\mu}^0 = 2 \sum_{f=1}^{N_f}  m_f \overline{q}_f
i \gamma_5 q_f + i N_f  2 q ,
\end{equation}
where $q$ is the local topological charge operator and is equal to 
$\frac{1}{16 \pi^2} G^\alpha_{\mu\nu} \tilde{G}^{\alpha\mu\nu}$ in the continuum. We put this identity between
the nucleon states and calculate the matrix element on the right-hand side with a 
momentum transfer $\vec{q}$  and take the $|\vec{q}| \rightarrow 0$ limit
\begin{eqnarray} \label{eq:awi}
\langle p^\prime s \left| A_\mu \right| p s \rangle s_\mu = \lim_{\vec{q} \rightarrow 0} \frac{i | \vec{s} |}
{\vec{q} \cdot \vec{s}}  \langle p^\prime,s  | 2 \sum_{f=1}^{N_f}  m_f \bar{q}_f i \gamma_5 q_f 
+ 2 i N_f q \, | p,s \rangle.
\end{eqnarray}
Lattice theory has finally accommodated vector chiral symmetry, the lack of which has hampered the 
development of chiral fermions on the lattice for many years. It is shown that when the lattice massless Dirac operator satisfies the Gingparg-Wilson relation $\gamma_5 D + D \gamma_5 = a D\gamma_5D$
with the overlap fermion being an explicit example~\cite{Neuberger:1997fp}, the
modified chiral transformation leaves the action invariant and gives rise to a chiral Jacobian factor
$J = e^{-2i\alpha Tr \gamma_5 (1 - \frac{1}{2}a D)}$ from the fermion determinant~\cite{Luscher:1998pqa}.
The index theorem~\cite{Hasenfratz:1998ri} shows that this Jacobian factor carries the correct chiral anomaly.
It is shown further that the local version of the overlap Dirac operator gives the topological
charge density operator in the continuum~\cite{Kikukawa:1998pd}, i.e.
\begin{equation}  \label{top_q}
Tr \gamma_5 (1 - \frac{1}{2}a D_{ov}(x,x) )= \frac{1}{16 \pi^2} G^\alpha_{\mu\nu} 
\tilde{G}^{\alpha\mu\nu}(x) + \mathcal{O}(a)
\end{equation}
Therefore, Eq.~(\ref{awi}) is exact on the lattice for the overlap fermion which gives the correct 
anomalous Ward identity at the continuum limit. Instead of calculating the matrix element of
the axial-vector current derived from the Noether procedure~\cite{Kikukawa:1998py,Hasenfratz:1998ri},
we shall calculate it from the r.h.s. of the AWI in Eq.~(\ref{awi}) through the form factors defined
in Eq.~(\ref{eq:awi}). 

In the lattice calculation with the overlap fermion, we note that the renormalization constant 
of the pseudoscalar density cancels that of the renormalization of the quark mass, i.e. 
$Z_m\,Z_P = 1$ for the chiral fermion. Also, the topological charge density, when calculated 
with the overlap Dirac operator as in the l.h.s of Eq.~(\ref{top_q}) is renormalized -- its integral
over the lattice volume is an integer satisfying the Atiya-Singer theorem.
Thus, when the matrix elements on the right-hand side of Eq.~(\ref{eq:awi}) are calculated with the overlap 
fermion and its Dirac operator, the flavor-singlet axial-vector current is automatically renormalized on the lattice
non-perturbatively \mbox{\it{\`{a} la}} anomalous Ward identity (AWI). 

Besides the fact that AWI admits non-perturbative renormalization on the lattice, 
the pseudoscalar density in DI and the topological density represent the low-frequency and
high-frequency parts of the divergence of the axial-vector quark loop respectively. It is learned that 
on the $24^3 \times 64$ lattice, a mere 20 pairs of the overlap low eigenmodes would
saturate more than 90\% of the pseudoscalar loop in configurations with 
zero modes~\cite{Gong:2013vja}. On the other hand, it is well-known that the
contribution to the triangle anomaly comes mainly from the cut-off of the regulator.
Therefore, the topological charge density represents the high-frequency contribution
of the axial-vector loop, albeit in a local form (the overlap operator is exponentially
local). Since the pseudoscalar density is totally dominated by the low modes, we expect
that the low-mode averaging (LMA) approach should be adequate for this term. To the extent that the signal for
the anomaly term is good, we should be able to calculate the flavor-singlet $g_A$ 
with the AWI . Both the overlap fermion for the quark loop and the overlap operator 
for the topological charge density are crucial in this approach.

\begin{figure}[hbt]
 %\vspace*{-2cm}
  \centering
\subfigure[]
 {{\includegraphics[width=0.48\hsize,angle=0]{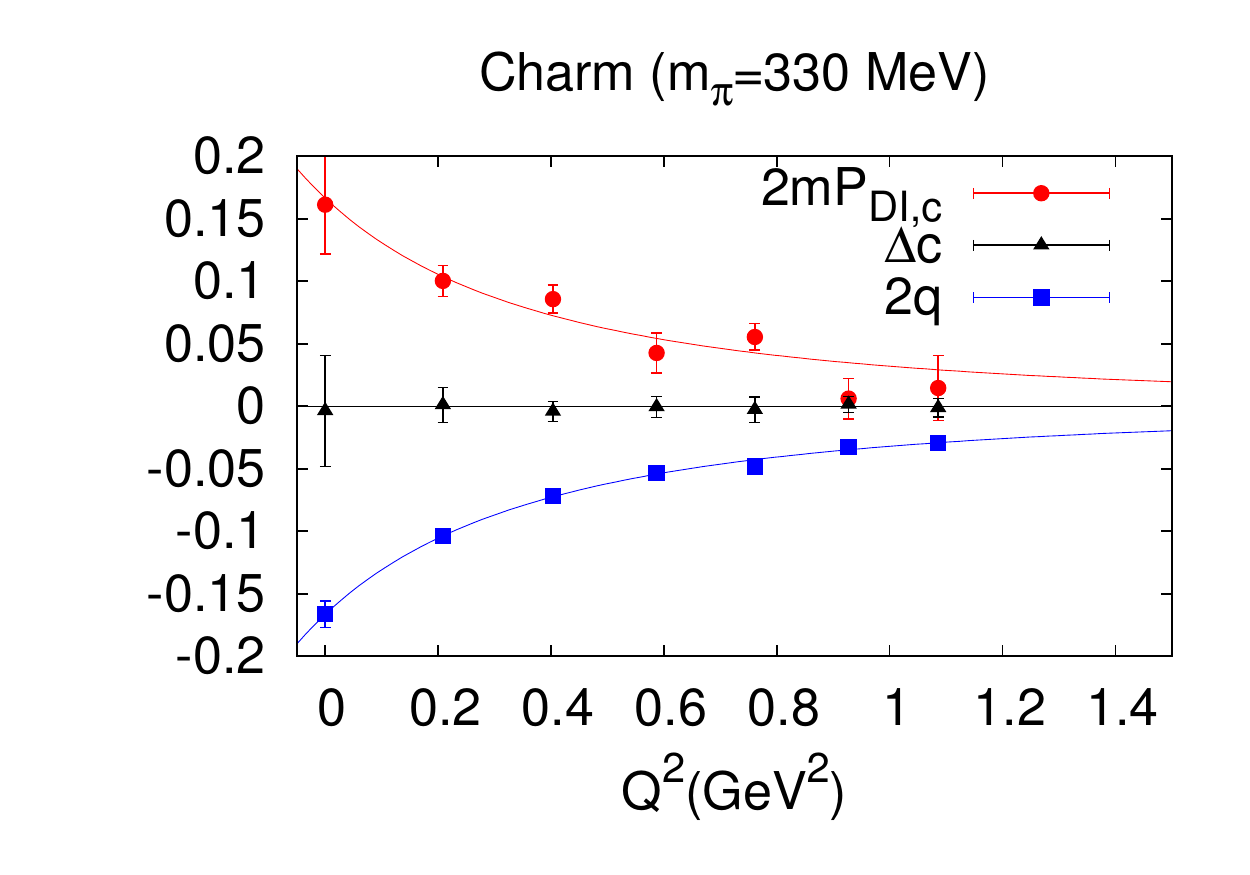}}
    \label{charm_spin}}
  \subfigure[]
  {{\includegraphics[width=0.48\hsize,angle=0]{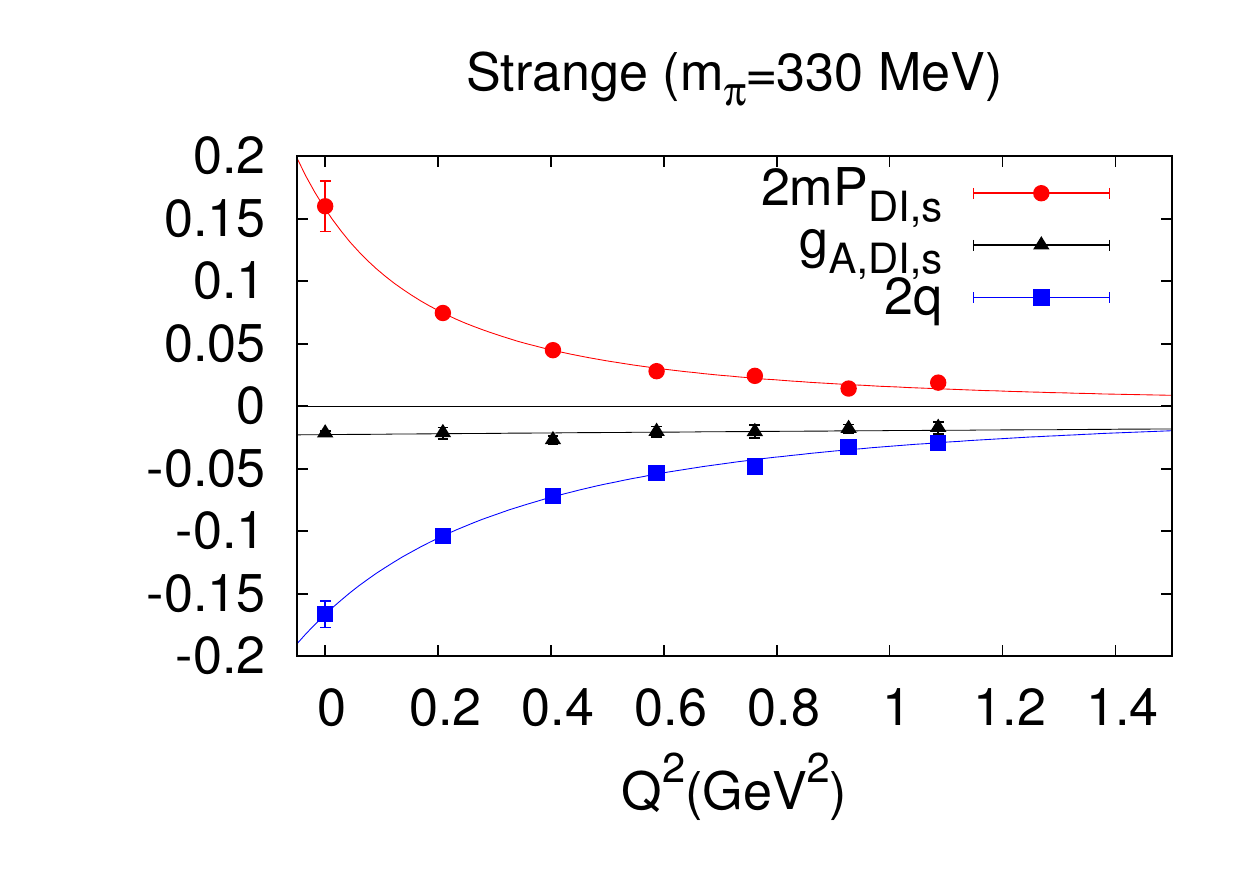}}
  \label{strange_spin}}  
  \caption{(a) The charm pseudoscalar and topological density contributions to the proton spin as
  a function of $Q^2$. (b) The same as in (a) for the strange.}
\end{figure}

With the approach described above, we have seen good signals on the
$24^3 \times 64$ lattice with the sea quark mass corresponding to a pion mass
at 330 MeV~\cite{Yang:2015xga}. 
We first show the results for the charm quark which contribute
only in the DI. The pseudoscalar density term and the topological charge density term
are plotted in Fig.~\ref{charm_spin} as a function of $Q^2$. We see that the
pseudoscalar term is large due to the large charm mass and positive, while the topological 
charge term is large and negative. When they are added together (black triangles in the
figure), it is consistent with zero for the whole range of $Q^2$. When extrapolated to
$Q^2 =0$, the charm gives zero contribution to the proton spin within error due to the cancellation 
between the pseudoscalar term and the topological term. It is shown~\cite{Franz:2000ee} 
that the leading term in the heavy quark expansion of the quark loop of the pseudoscalar density, 
i.e. $2mP$ is the topological charge $\frac{2i}{16 \pi^2}tr_c G_{\mu\nu} \tilde{G}_{\mu\nu}$, 
but with a negative sign. Thus, one expects that there is no contribution to the quark spin from heavy quarks
to leading order. It  appears that the charm quark is heavy enough so that the $\mathcal{O}(1/m^2)$ correction
is small. We take this as a cross check of the validity of our numerical estimate
of the DI calculation of the quark loop as well as the anomaly contribution.

The contributions from the strange are also calculated and shown in Fig.~\ref{strange_spin}. The $2mP$ contribution
is slightly smaller than that of $2q$ and results in a net small negative value for the sum of $2mp$ and
$2q$ at finite $Q^2$. After a dipole fit, we obtain $\Delta s = -0.026(5)$ at $m_{\pi} = 330$ MeV. Here, 
$\Delta s$ denotes the contributions for both $s$ and $\bar{s}$. $\Delta u$ and $\Delta d$ are similarly
defined in the following.  
 
 \begin{figure}[hbt]
%\vspace*{-1cm}
  \centering
 \subfigure[] 
 {\includegraphics[width=0.48\hsize,angle=0]{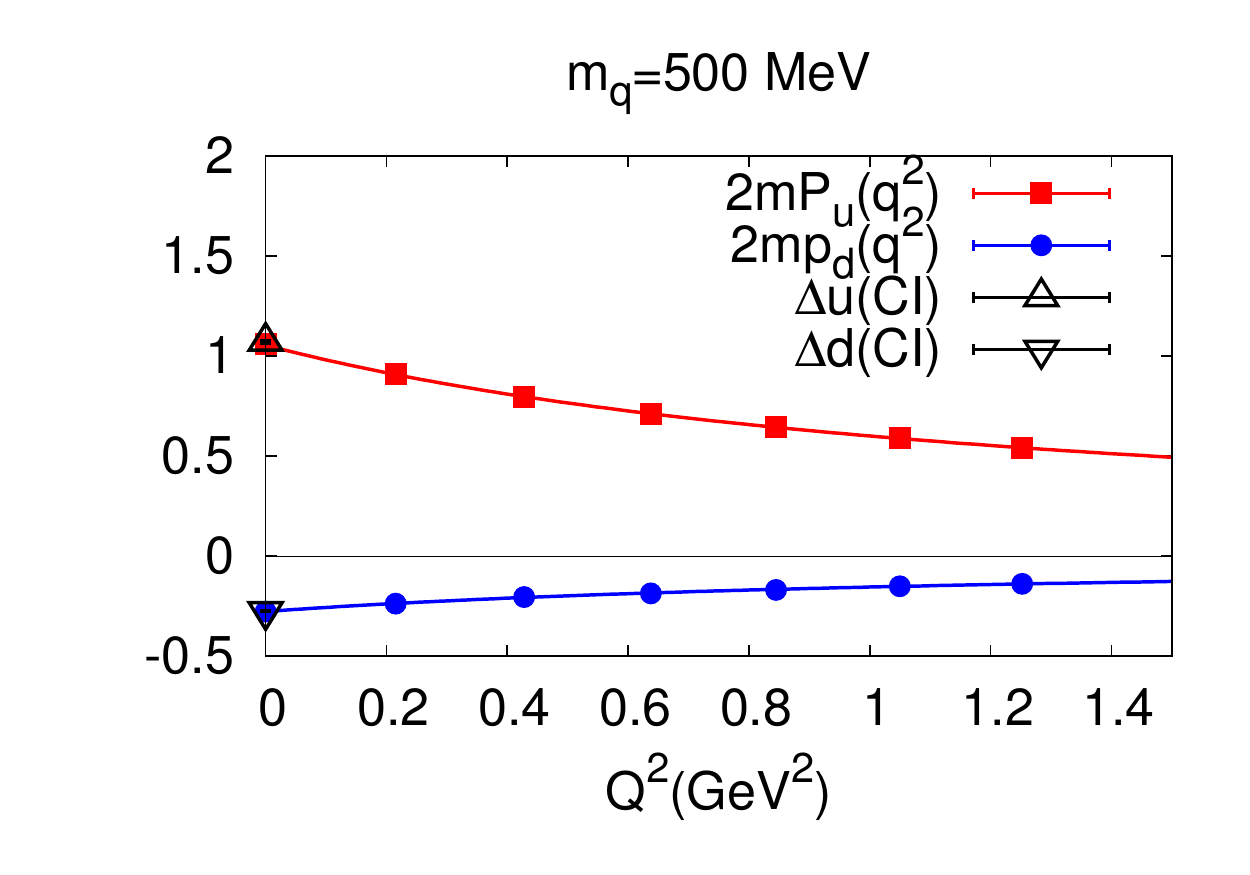}}
 \subfigure[]
 {\includegraphics[width=0.48\hsize,angle=0]{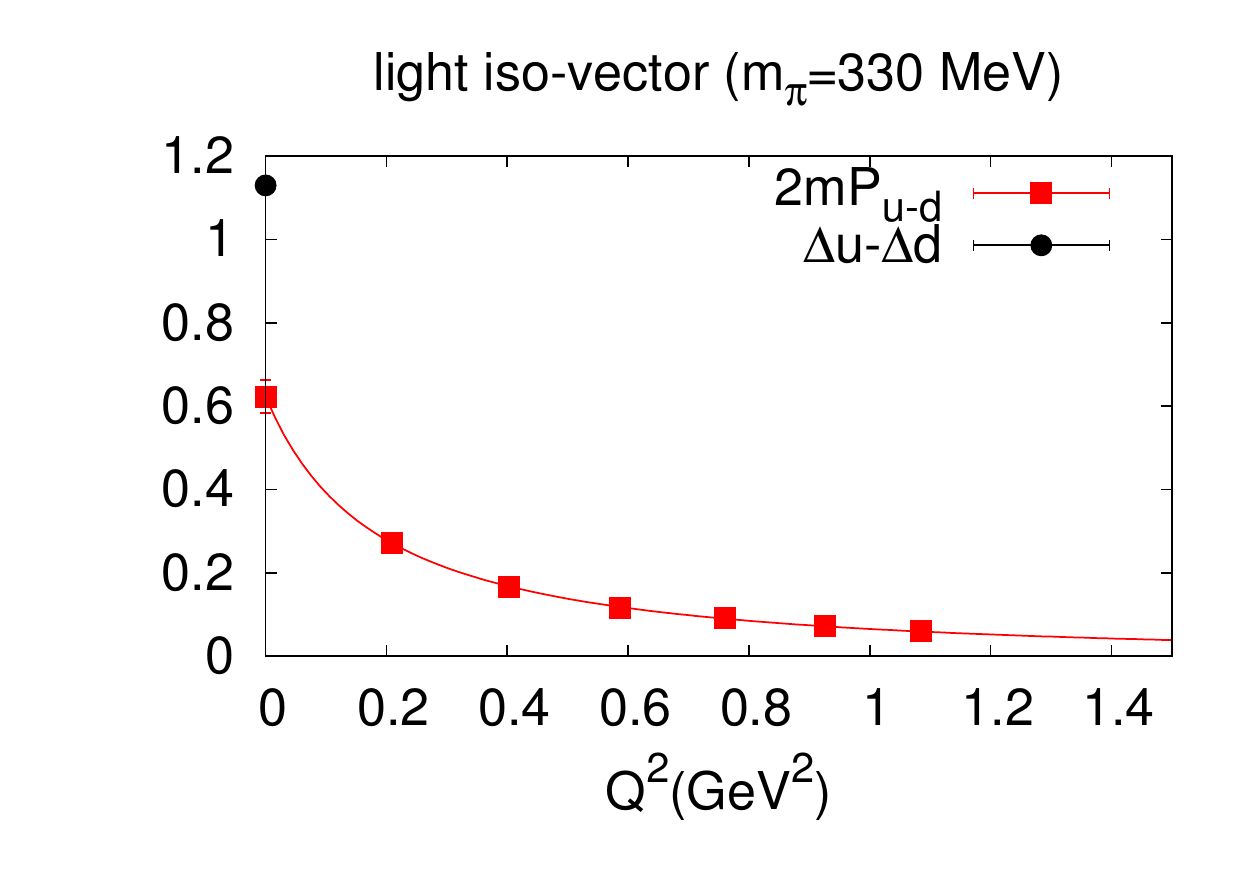}} 
  \caption{(a) The quark spin of the proton-like baryon with $m_q\sim500$ MeV from both the axial vector current and
the pseudoscalar term through AWI. In this case, the DI contribution of $2mP$ is canceled by the topological charge term. (b) The same as in (a) for light quarks at the unitary point for isovector $g_A^3$ which involves only CI.}
  \label{fig:heavy}
\end{figure}

Since this $\Delta s$ is quite a bit smaller than the experimental value, we explore the possible finite volume effect and 
the fact that the induced pseudoscalar form factor $h_A(q^2)$ has been neglected in the $Q^2$ extrapolation which does not contribute at the $Q^2 = 0$ limit as in Eq.~(\ref{eq:awi}), but has a contribution at finite $Q^2$~\cite{Liu:1995kb}. We shall check this in the connected insertion (CI) calculation. As can be seen in Fig.~\ref{fig:heavy} for  $m_q\sim500$ MeV, both $\Delta u$ and $\Delta d$ in CI calculated from the axial-vector current and
renormalized with $Z_A$ from the isovector Ward identity are well reproduced through the $Q^2$ extrapolation  of
$2mP$ with a dipole form.  Whereas, in the case of light quarks at the unitary point, $g_A^3 = 1.13(2)$ from the axial-vector current is $1.8(1)$ times larger than $0.62(4)$ from the dipole extrapolation of $2mP$. This is most likely due to the ignorance of the induced pseudoscalar form factor $h_A(q^2)$ as well as the finite volume effect at small $Q^2$ which is well known to plague the $Q^2$ extrapolation of the nucleon magnetic form factor.

 \begin{figure}[htb]
% \vspace*{-1cm}
 \centering
% \hspace{2cm}
  {\includegraphics[width=0.7\hsize,angle=0]{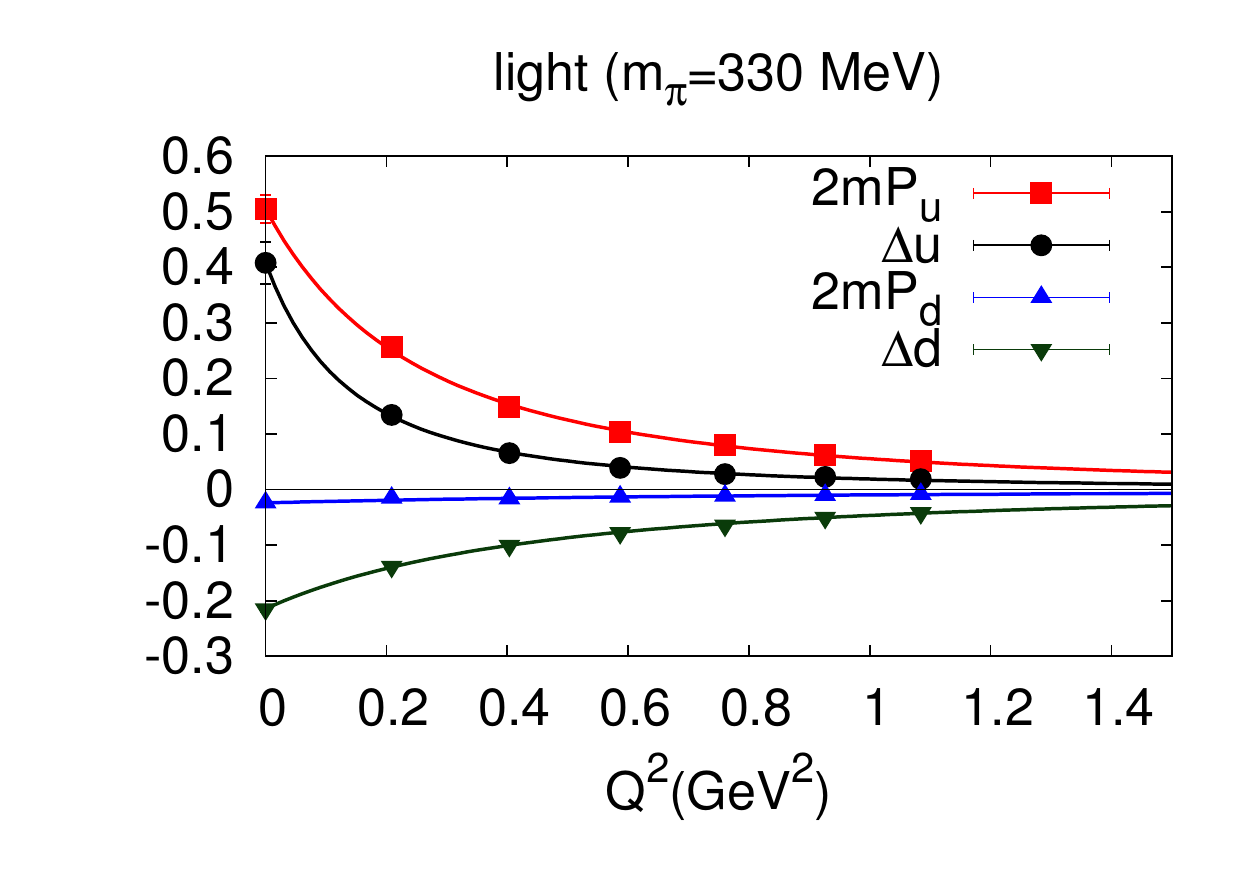}\label{ud_spin}
\caption{The combined pseudoscalar contribution from both the connected insertion (CI) and DI ($2mP_{u/d}$ in the plot), along with the overall quark spin from both pseudoscalar and topological charge ($g_{A,d}$). The plot corresponds to the unitary point with $m_{\pi} = 330$ MeV.}}
\end{figure}

At the unitary point, when the valence $u/d$ mass matches that of the light sea, Fig.~\ref{ud_spin} shows the quark spin contribution from the combined pseudoscalar terms $2mP_{u/d}$ of the  CI and DI with a dipole extrapolation. Also plotted are the overall quark spin $\Delta u/\Delta d$ by including the topological charge contribution.  In this case, we obtain $\Delta u + \Delta d = 0.19(3)$ and $\Delta u - \Delta d = 0.62(4)$ at $Q^2 = 0$ from a dipole extrapolation in $Q^2$. As we discussed above, the fact that $g_A^3$ from the axial current is
$1.8(1)$ times larger than that of $\Delta u - \Delta d$ through the Ward identity approach is most likely due to the neglect of the induced pseudoscalar form factor $h_A(q^2)$ and the finite volume effect in the $Q^2$ extrapolation. We apply this  $1.8(1)$ factor as an estimate to correct the present AWI approach and obtain 
$\Delta u + \Delta d = 0.35(6)$, $\Delta s = -0.05(1)$. Thus the total estimated spin $\Delta \Sigma= 0.30(6)$ at the unitary point is consistent with the present experimental results which are between 0.2 and 0.3. We expect that, at lighter quark masses, $\Delta \Sigma$ will be smaller.
    
The above results are from the $24^3 \times 64$ lattice at $m_{\pi} = 330$ MeV with 200 configurations.
The nucleon propagator in the DI has been calculated with the smeared-grid noise
source with time dilution which covers all time slices in order to have reasonable
statistics for the DI.

\section{Glue spin}

      It has been pointed out that decomposing glue angular momentum into glue spin and orbital angular momentum
is only feasible in a specific gauge~\cite{Jaffe:1989jz}. Making contact with the parton picture, a spin sum rule involving
quark and glue spins and orbital angular momenta is derived in the light-cone gauge (i.e. $A^+ =0$) with nucleon in the infinite momentum frame~\cite{Jaffe:1989jz}.  The longitudinal glue spin content is
\be   \label{jm89}
S_G^3 = \langle p, s|\int d^3 x Tr (\vec{E} \times \vec{A})^3 |p, s\rangle/\langle p, s| p, s\rangle,
\ee
where the nucleon state is in the infinite momentum frame and the gauge potential and the gauge field are in the
light-cone gauge. Similarly, a gauge-invariant glue helicity distribution is defined with the light-cone 
correlation function~\cite{Manohar:1990kr} 
\begin{equation}   \label{manohar}
  \Delta g (x) S^+ = \frac{i}{2xP^+} \int
  \frac{d\xi^-}{2\pi} e^{-ixP^+\xi^-} \langle PS| F^{+\alpha}_a(\xi^-) {\cal L}^{ab}(\xi^-,0)\tilde F_{\alpha,b}^{~+} (0)|PS\rangle \ ,
\end{equation}
where \mbox{$\tilde F^{\alpha\beta} = (1/2)\  \epsilon^{\alpha\beta\mu\nu}F_{\mu\nu}$} is
in the adjoint representation with $\mathcal{A}^+ \equiv T^cA^+_c$, so is  the light-cone link 
\mbox{${\cal L}(\xi^- ,0) = P\exp[-ig\int^{\xi^-}_0  \mathcal{A}^+(\eta^-,0_\perp)\ d\eta^-]$} . 

Since lattice QCD is formulated in Euclidean time, it is not equipped to address the light-cone gauge or the
light-cone coordinates and; as such, one is not able to calculate $\Delta G$ as defined in Eqs.~(\ref{jm89})
and (\ref{manohar}) on the lattice directly. 

On the other hand, a gauge-invariant decomposing of the proton spin has been 
formulated~\cite{Chen:2008ag,Chen:2009mr} and examined in various 
contexts~\cite{Wakamatsu:2010qj,Hatta:2011zs,Cho:2010cw,Leader:2013jra}. It is
based on the canonical energy momentum tensor, instead of that in the symmetric Belinfonte form. 
The glue spin operator is
\bea\la{ExA}
\vec{S}_g =  \vec{E}^a\times\vec{A}^a_{phys}
\eea
where $A_{\mu\, phys}$ is the physical component of 
the gauge field $A_{\mu}$ which is decomposed into $A_{\mu\,phys}$ and a pure gauge part
as in QED,
\bea\la{Amu}
A_{\mu} = A_{\mu\,phys} + A_{\mu\,pure}.
\eea
They transform homogeneously and inhomogeneously with respect to gauge transformation respectively,
\bea\label{eq5}
A_{\mu\,phys}  &&  \rightarrow A'_{\mu\,phys}= g A_{\mu\,phys}g^{-1} \nonumber \\
A_{\mu\,pure} &&  \rightarrow A'_{\mu\,pure} = gA_{\mu\,pure}g^{-1}-\frac{i}{g_0}g\partial_\mu g^{-1}, 
\eea
where $g$ is the gauge transformation matrix and $g_0$ is the coupling constant. In oder to have a unique solution, conditions are set as follows: the pure gauge part does not give rise to a field tensor by itself and $A_{phys\,\mu}$ satisfies the non-Abelian Coulomb gauge condition
\bea
F_{\mu\nu\,pure} &&=\partial_\mu A_{\nu\,pure} -\partial_{\nu}A_{\mu\,pure}- ig_0[A_{\mu\,pure},\, 
A_{\nu\,pure}]=0 \nonumber \\
D_i A_{i\,phys} &&= \partial_i A_{i\,phys}-ig_0[A_i,\,A_{i\,phys}] =0. 
\eea
This is analogous to the the situation in QED where the photon spin and orbital angular
momentum can be defined~\cite{cdg89,en94,Bliokh:2012zr,Bliokh:2014ara} from the canonical energy-momentum 
tensor 
\begin{eqnarray}   \label{QED-SL}
\mbox{\boldmath $S$}_A &=& \int \,
 \mbox{\boldmath $E_{\perp}$} \times 
 \mbox{\boldmath $A$}_{\perp} \,d^3 x , \\
 \mbox{\boldmath $L$}_A &=& \sum_i \int \,E_i^{\perp} \,
 (\mbox{\boldmath $x$} \times \bm{\nabla}) \,  A_i^{\perp} \,d^3 x ,
\end{eqnarray}
where $\perp$ denotes the transverse part. Since they are defined in terms of the transverse parts,
they are gauge invariant. However, this gauge invariant definition breaks Lorentz invariance. 
Nevertheless, it is shown that the `spin' and `orbital' angular momentum so defined are conserved for
a free field~\cite{en94}. Furthermore, they are observables and can be measured in experiments
through interaction with matter. In 1936, Beth had observed one component of the spin angular momentum
of light~\cite{beth36}, by measuring the tongue on a birefringent plate exerted by a circularly polarized light. 
Also, it is shown~\cite{bsw92} that the orbital angular momentum of a paraxial laser beam can be measured. 
Even though gauge invariance is preserved in this canonical formulation, the spin and orbital AM operators are not boost invariant. Since the experiments are conducted in the lab, the formulation is adequate for this single reference frame. 

After integrating the longitudinal momentum $x$, the light-cone operator for the matrix element has the following expression for the glue helicity~\cite{Hatta:2011zs,Ji:2013fga} 
\bea\la{eq4}
H_g &=& \Bigg[ \vec{E}^a(0)\times (\vec{A}^a(0)-\frac{1}{\nabla^+} (\vec{\nabla}A^{+,b})\mathcal{L}^{ba}(\xi^-,0))\Bigg]^z
\eea
It is recently shown~\cite{Ji:2013fga} that when boosting the glue spin density operator $\vec{S}_g$ in Eq.(\ref{ExA}) to the infinite momentum frame (IMF), the second term in the parentheses on the right side of
Eq. (\ref{eq4}) is $\vec{A}_{pure}$. Thus $H_g$ is the glue spin density operator $\vec{S}_g$ in
the IMF along the direction of the moving frame. In other words, the longitudinal glue spin operator turns
into the helicity operator in the IMF.

To carry out a lattice calculation of the matrix element of the glue spin operator, it is realized~\cite{yl14} that 
$A_{\mu\, phys}$ is related to that fixed in the Coulomb gauge, i.e. $A_{\mu\, phys} = g_c^{-1} A_c g_c$ where 
$A_c$ is the gauge potential fixed to the Coulomb gauge and $g_c$ is the gauge transformation that fixes the Coulomb gauge. Since $\vec{S}_g$ is traced over color, the spin operator is then
\bea\label{eq7}
\vec{S}_G = \int d^3 x\, Tr (g_c\vec{E}g_c^{-1}\times\vec{A}_c) =  \int d^3 x\, Tr (\vec{E}_c\times
\vec{A}_c)
 \eea
where $\vec{E}_c$ is the electric field in the Coulomb gauge. Although it is gauge invariant since both
$E$ and $A_{phys}$ transform homogeneouly, it is frame dependent and thus depends on the proton momentum.
Its IMF value corresponds to $\Delta G$ which is measurable experimentally from high energy proton-proton
scattering. The important outcome of the derivation is that glue spin content is amenable to lattice QCD
calculation. To the extent that it can be calculated at large enough momentum frame of the proton with
enough precision, it can be compared to the experimental glue helicity $\Delta G$.

    The first attempt to calculate $S_G$ on the lattice has been carried out on the same set of $2+1$ flavor dynamical
domain-wall configurations on the $24^3 \times 64$ lattice with the sea pion mass at 330 MeV~\cite{Sufian:2014jma}. The electric field $\vec{E}$ is constructed from the overlap Dirac operator defined in Eq.~(\ref{Glue_Tensor}).
 The gauge potential $\vec{A}$ is obtained from the unsmeared gauge link. We obtained results for the longitudinal nucleon momenta $p_z = n (2\pi/La)$ with $n=0, 1, 2$ which correspond to 0, 460 MeV and 920 MeV and for the case of quark masses in the nucleon propagator which correspond to $m_{\pi} = 380$ MeV and 640 MeV. The unrenormalized results are presented in Fig.~\ref{glue_spin}.
\begin{figure}
\begin{center}
    \includegraphics[width=.80\textwidth]{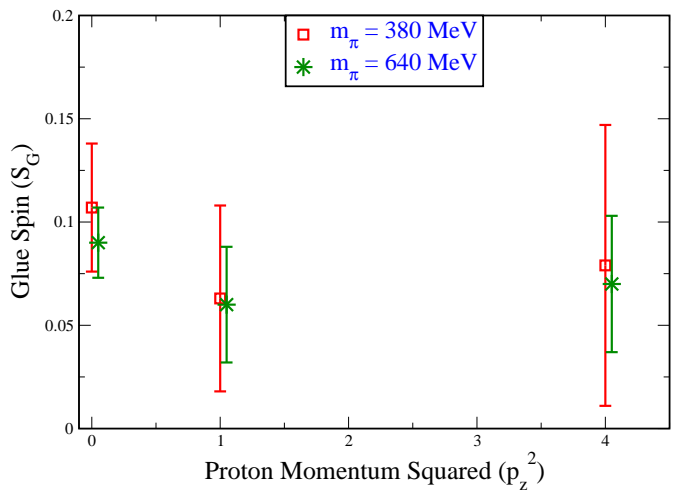}
\end{center}
\caption{The results of glue spin $S_G$ in longitudinally polarized proton with longitudinal momenta
at 0, 460 MeV and 930 MeV. The quark masses in the nucleon propagator correspond to $m_{\pi} = 380$ and
640 MeV.}
\la{glue_spin}
\end{figure} 
We see that the preliminary results in Fig.~\ref{glue_spin} are quite noisy and, as a result, one cannot discern
the $p_z$ behavior. The signal can be improved by smearing the link, but it is a challenge to reach large $p_z$
on the lattice. Since one needs $p_z$ to be less than the cutoff, i.e. $p_z a \ll 1$ to avoid large discretization error,
this will require a large lattice size $L$ so that $m_{\pi}La > 6$ for the nucleon. We note that the question how large a $p_z$ is needed to have the quasi PDFs coincide with the PDFs has been studied in a spectator diquark 
model~\cite{Gamberg:2014zwa}. It is found that it is necessary to have $p_z$ as large as 4 GeV for the
quasi PDFs to be a good approximation of the PDFs.

\section{Summary}
 We have reported the current lattice efforts in calculating the quark spin, quark orbital angular momentum, glue angular momentum and glue spin in the nucleon. A complete decomposition of the proton momentum and spin into its quark and glue components is  given in a quenched approximation. In this case, the glue angular momentum  is not further divided  into spin and orbital angular momentum parts. The quark spin calculation is recently carried out from the anomalous Ward identity with chiral fermions and is found to be small mainly due to the large negative anomaly term which is believed to be the culprit of the `proton spin crisis'. An exploratory lattice calculation of $S_G$ in the non-Abelian Coulomb gauge is carried out~\cite{Sufian:2014jma} which has large errors and the nucleon momentum is limited to $\sim 1$ GeV. The signal of the glue spin $S_G$ can be improved with smearing, but the major challenge is to have a lattice with fine enough lattice spacing to accommodate large momentum states and show that the infinite momentum extrapolation can be made under control.

\section*{Acknowledgments}

This work is partially supported by USDOE grant DE-FG05-84ER40154. The author would like to thank 
X.S. Chen, X. Ji, L. Gamberg, Y. Hatta, E. Leader, C.Lorc\'{e}, M. Wakamatsu, and Y. Zhao for helpful and insightful discussions. He also thanks E. Leader and D. Stamenov for providing the strange quark spin contribution from their analysis.

%\begin{thebibliography}{000} %for 3 digits
%\begin{thebibliography}{00}  %for 2 digits


\begin{thebibliography}{0}    %for 1 digit

  % 
\bibitem{deFlorian:2009vb} 
  D.~de Florian, R.~Sassot, M.~Stratmann and W.~Vogelsang,
  % ``Extraction of Spin-Dependent Parton Densities and Their Uncertainties,''
  Phys.\ Rev.\ D {\bf 80}, 034030 (2009)
  [arXiv:0904.3821 [hep-ph]].
  % 
\bibitem{Dong:1995rx}
  S.~J.~Dong, J.~-F.~Lagae, K.~F.~Liu,
  % ``Flavor singlet g(A) from lattice QCD,''
  Phys.\ Rev.\ Lett.\  {\bf 75}, 2096-2099 (1995), 
  [hep-ph/9502334].
  % 
  % 
\bibitem{Fukugita:1994fh} 
  M.~Fukugita, Y.~Kuramashi, M.~Okawa and A.~Ukawa,
  % ``Proton spin structure from lattice QCD,''
  Phys.\ Rev.\ Lett.\  {\bf 75}, 2092 (1995), 
  [hep-lat/9501010].
  % 
  % 
\bibitem{Gusken:1999as} 
  S.~Gusken {\it et al.}  [TXL Collaboration],
  % ``Flavor singlet axial vector coupling of the proton with dynamical Wilson fermions,''
  Phys.\ Rev.\ D {\bf 59}, 114502 (1999).
  % 
  % 
\bibitem{QCDSF:2011aa} 
  G.~S.~Bali {\it et al.}  [QCDSF Collaboration],
  % ``Strangeness Contribution to the Proton Spin from Lattice QCD,''
  Phys.\ Rev.\ Lett.\  {\bf 108}, 222001 (2012), 
  [arXiv:1112.3354 [hep-lat]].
  % 
  %
  \bibitem{Engelhardt:2012gd} 
  M.~Engelhardt,
  %``Strange quark contributions to nucleon mass and spin from lattice QCD,''
  Phys.\ Rev.\ D {\bf 86}, 114510 (2012)
  [arXiv:1210.0025 [hep-lat]].
   % 
   %
\bibitem{Abdel-Rehim:2013wlz} 
  A.~Abdel-Rehim, C.~Alexandrou, M.~Constantinou, V.~Drach, K.~Hadjiyiannakou, K.~Jansen, G.~Koutsou and A.~Vaquero,
  % ``Disconnected quark loop contributions to nucleon observables in lattice QCD,''
  arXiv:1310.6339 [hep-lat].
  % 
  % 
\bibitem{Babich:2010at} 
  R.~Babich, R.~C.~Brower, M.~A.~Clark, G.~T.~Fleming, J.~C.~Osborn, C.~Rebbi and D.~Schaich,
  % ``Exploring strange nucleon form factors on the lattice,''
  Phys.\ Rev.\ D {\bf 85}, 054510 (2012)
  [arXiv:1012.0562 [hep-lat]].
  %
   %
   \bibitem{Leader:2014uua} 
  E.~Leader, A.~V.~Sidorov and D.~B.~Stamenov,
  %``New analysis concerning the strange quark polarization puzzle,''
  Phys.\ Rev.\ D {\bf 91}, no. 5, 054017 (2015)
  [arXiv:1410.1657 [hep-ph]].
  %
  %
  \bibitem{leader15}
 E.~Leader and D.~B.~Stamenov, private communication.
  %
 %
\bibitem{Yang:2015xga} 
  Y.~B.~Yang, M.~Gong, K.~F.~Liu and M.~Sun,
  %``Quark Spin in Proton from Anomalous Ward Indentity,''
  PoS LATTICE {\bf 2014}, 138 (2014)
  [arXiv:1504.04052 [hep-ph]].
  %
  % 
\bibitem{Mathur:1999uf}
  N.~Mathur, S.~J.~Dong, K.~F.~Liu, L.~Mankiewicz, N.~C.~Mukhopadhyay,
  % ``Quark orbital angular momentum from lattice QCD,''
  Phys.\ Rev.\  {\bf D62}, 114504 (2000),   [hep-ph/9912289].
  % 
  % 
\bibitem{Hagler:2003jd}
  P.~Hagler {\it et al.} [LHPC and SESAM Collaborations],
  % ``Moments of nucleon generalized parton distributions in lattice QCD,''
  Phys.\ Rev.\  {\bf D68}, 034505 (2003), 
  [hep-lat/0304018].
  % 
  % 
\bibitem{Gockeler:2003jfa} 
  M.~Gockeler {\it et al.}  [QCDSF Collaboration],
  % ``Generalized parton distributions from lattice QCD,''
  Phys.\ Rev.\ Lett.\  {\bf 92}, 042002 (2004)
  [hep-ph/0304249].
  % 
   % 
\bibitem{Brommel:2007sb} 
  D.~Brommel {\it et al.}  [QCDSF-UKQCD Collaboration],
  % ``Moments of generalized parton distributions and quark angular momentum of the nucleon,''
  PoS LATTICE {\bf 2007}, 158 (2007), 
  [arXiv:0710.1534 [hep-lat]].
  %    % 
\bibitem{Bratt:2010jn} 
  J.~D.~Bratt {\it et al.}  [LHPC Collaboration],
  % ``Nucleon structure from mixed action calculations using 2+1 flavors of asqtad sea and domain 
  % wall valence fermions,''
  Phys.\ Rev.\ D {\bf 82}, 094502 (2010)
  [arXiv:1001.3620 [hep-lat]].
  % 
   %
\bibitem{Alexandrou:2011nr} 
  C.~Alexandrou, J.~Carbonell, M.~Constantinou, P.~A.~Harraud, P.~Guichon, K.~Jansen, C.~Kallidonis and T.~Korzec {\it et al.},
  % ``Moments of nucleon generalized parton distributions from lattice QCD,''
  Phys.\ Rev.\ D {\bf 83}, 114513 (2011)
  [arXiv:1104.1600 [hep-lat]].
  %
 %
  \bibitem{Syritsyn:2011vk}
  S.~N.~Syritsyn, J.~R.~Green, J.~W.~Negele, A.~V.~Pochinsky, M.~Engelhardt, P.~Hagler, B.~Musch and W.~Schroers,
  %``Quark Contributions to Nucleon Momentum and Spin from Domain Wall fermion calculations,''
  PoS LATTICE {\bf 2011} (2011) 178
  [arXiv:1111.0718 [hep-lat]].  
  %
   %  
\bibitem{Alexandrou:2013joa} 
  C.~Alexandrou, M.~Constantinou, S.~Dinter, V.~Drach, K.~Jansen, C.~Kallidonis and G.~Koutsou,
  % ``Nucleon form factors and moments of generalized parton distributions using $N_f=2+1+1$ twisted mass %fermions,''
  Phys.\ Rev.\ D {\bf 88}, 014509 (2013)
  [arXiv:1303.5979 [hep-lat]].
  % 
   %
\bibitem{Adolph:2012ca} 
  C.~Adolph {\it et al.}  [COMPASS Collaboration],
  % ``Leading and Next-to-Leading Order Gluon Polarization in the Nucleon and Longitudinal Double Spin Asymmetries from Open Charm Muoproduction,''
  Phys.\ Rev.\ D {\bf 87}, 052018 (2013); 
  [arXiv:1211.6849 [hep-ex]];
  C.~Adolph {\it et al.}  [COMPASS Collaboration],
  % ``Leading order determination of the gluon polarisation from DIS events with high-$p_T$ hadron pairs,''
  Phys.\ Lett.\ B {\bf 718}, 922 (2013)
  [arXiv:1202.4064 [hep-ex]].
  % 
  % 
\bibitem{Djawotho:2011zz} 
  P.~Djawotho [STAR Collaboration],
  % ``Gluon polarization and jet production at STAR,''
  J.\ Phys.\ Conf.\ Ser.\  {\bf 295}, 012061 (2011).
  % 
  % 
\bibitem{Airapetian:2010ac} 
  A.~Airapetian {\it et al.}  [HERMES Collaboration],
  % ``Leading-Order Determination of the Gluon Polarization from high-p(T) Hadron Electroproduction,''
  JHEP {\bf 1008}, 130 (2010)
  [arXiv:1002.3921 [hep-ex]].
  % 
  % 
\bibitem{Stolarski:2010zz} 
  M.~Stolarski [COMPASS Collaboration],
  % ``The COMPASS results on longitudinal spin effects and future measurements,''
  Nucl.\ Phys.\ Proc.\ Suppl.\  {\bf 207-208}, 53 (2010).
  % 
  % 
\bibitem{Adare:2008qb} 
  A.~Adare {\it et al.}  [PHENIX Collaboration],
  %``Inclusive cross section and double helicity asymmetry for pi^0 production in $p^+ p$ collisions at $\sqrt{s}=62.4$ GeV,''
  Phys.\ Rev.\ D {\bf 79}, 012003 (2009)
  [arXiv:0810.0701 [hep-ex]].
  % 
  %
\bibitem{Aschenauer:2013woa} 
  E.~C.~Aschenauer, A.~Bazilevsky, K.~Boyle, K.~O.~Eyser, R.~Fatemi, 
  C.~Gagliardi, M.~Grosse-Perdekamp and J.~Lajoie {\it et al.},
  % ``The RHIC Spin Program: Achievements and Future Opportunities,''
  arXiv:1304.0079 [nucl-ex].
  % 
  % 
\bibitem{Alekseev:2010hc} 
  M.~G.~Alekseev {\it et al.}  [COMPASS Collaboration],
  % ``The Spin-dependent Structure Function of the Proton g_1^p and a Test of the Bjorken Sum Rule,''
  Phys.\ Lett.\ B {\bf 690}, 466 (2010)
  [arXiv:1001.4654 [hep-ex]]; 
  Phys.\ Lett.\ B {\bf 693}, 227 (2010)
  [arXiv:1007.4061 [hep-ex]].
  % 
  % 
  \bibitem{deFlorian:2014yva} 
  D.~de Florian, R.~Sassot, M.~Stratmann and W.~Vogelsang,
  %``Evidence for polarization of gluons in the proton,''
  Phys.\ Rev.\ Lett.\  {\bf 113}, no. 1, 012001 (2014)
  [arXiv:1404.4293 [hep-ph]].  
 % 
 %
 \bibitem{Adamczyk:2014ozi} 
  L.~Adamczyk {\it et al.}  [STAR Collaboration],
  %``Precision Measurement of the Longitudinal Double-spin Asymmetry for Inclusive Jet 
  %Production in Polarized Proton Collisions at $\sqrt{s}=200$ GeV,''
  arXiv:1405.5134 [hep-ex].
  %
  %
  \bibitem{Adare:2014hsq} 
  A.~Adare {\it et al.}  [PHENIX Collaboration],
  %``Inclusive double-helicity asymmetries in neutral-pion and eta-meson production 
  % in $\vec{p}+\vec{p}$ collisions at $\sqrt{s}=200$ GeV,''
  Phys.\ Rev.\ D {\bf 90}, no. 1, 012007 (2014)
  [arXiv:1402.6296 [hep-ex]].
  %
  %
  \bibitem{Brodsky:2006ha} 
   S.~J.~Brodsky and S.~Gardner,
  % ``Evidence for the Absence of Gluon Orbital Angular Momentum in the Nucleon,''
  Phys.\ Lett.\ B {\bf 643}, 22 (2006)
  [hep-ph/0608219].
  % 
  %
\bibitem{Liu:2012nz}
  K.~F.~Liu, M.~Deka, T.~Doi, Y.~B.~Yang, B.~Chakraborty, Y.~Chen, S.~J.~Dong and T.~Draper {\it et al.},
  %``Quark and Glue Momenta and Angular Momenta in the Proton --- a Lattice Calculation,''
  PoS LATTICE {\bf 2011} (2011) 164
  [arXiv:1203.6388 [hep-ph]].
%
%
\bibitem{Deka:2013zha} 
  M.~Deka, T.~Doi, Y.~B.~Yang, B.~Chakraborty, S.~J.~Dong, T.~Draper, M.~Glatzmaier, M.~Gong, 
  H.W.~Lin, K.F.~Liu, D.~Mankame, N.~Mathur, and T.~Streuer,
  %``Lattice study of quark and glue momenta and angular momenta in the nucleon,''
  Phys.\ Rev.\ D {\bf 91}, no. 1, 014505 (2015)
  [arXiv:1312.4816 [hep-lat]].
%
%
\bibitem{Ji:1996ek} 
  X.~D.~Ji,
  %``Gauge-Invariant Decomposition of Nucleon Spin,''
  Phys.\ Rev.\ Lett.\  {\bf 78}, 610 (1997)
  [hep-ph/9603249].
  %
  %
  %
%
 \bibitem{Wilcox:2002zt} 
  W.~Wilcox,
  %``Continuum moment equations on the lattice,''
  Phys.\ Rev.\ D {\bf 66}, 017502 (2002)
  [hep-lat/0204024].
  %
  %
\bibitem{Liu:2007hq} 
  K.~F.~Liu, A.~Alexandru and I.~Horvath,
  %``Gauge field strength tensor from the overlap Dirac operator,''
  Phys.\ Lett.\ B {\bf 659}, 773 (2008)
  [hep-lat/0703010 [HEP-LAT]].
%
%
\bibitem{Alexandru:2008fu} 
  A.~Alexandru, I.~Horvath and K.~F.~Liu,
  %``Classical Limits of Scalar and Tensor Gauge Operators Based on the Overlap Dirac Matrix,''
  Phys.\ Rev.\ D {\bf 78}, 085002 (2008)
  [arXiv:0803.2744 [hep-lat]].
%
%
\bibitem{Teryaev:1999su} 
  O.~V.~Teryaev,
  %``Spin structure of nucleon and equivalence principle,''
  hep-ph/9904376.
  % 
%
\bibitem{Brodsky:2000ii} 
  S.~J.~Brodsky, D.~S.~Hwang, B.~-Q.~Ma and I.~Schmidt,
  %``Light cone representation of the spin and orbital angular momentum of relativistic 
  %  composite systems,''
  Nucl.\ Phys.\ B {\bf 593}, 311 (2001)
  [hep-th/0003082].
  % 
  %
\bibitem{Glatzmaier:2014sya} 
  M.~Glatzmaier and K.~F.~Liu,
  %``Perturbative Renormalization and Mixing of Quark and Glue Energy-Momentum 
  % Tensors on the Lattice,''
  arXiv:1403.7211 [hep-lat].
%
%
\bibitem{Liu:1993cv}
  K.~F.~Liu, S.~J.~Dong,
  % ``Origin of difference between anti-d and anti-u partons in the nucleon,''
  Phys.\ Rev.\ Lett.\  {\bf 72}, 1790-1793 (1994), 
  [hep-ph/9306299].
  % 
%
\bibitem{Liu:1998um}
  K.~F.~Liu, S.~J.~Dong, T.~Draper, D.~Leinweber, J.~H.~Sloan, W.~Wilcox, R.~M.~Woloshyn,
  %``Valence QCD: Connecting QCD to the quark model,''
  Phys.\ Rev.\  {\bf D59}, 112001 (1999), 
  [hep-ph/9806491].
  % 
  % 
\bibitem{Liu:1999ak}
  K.~F.~Liu,
  % ``Parton degrees of freedom from the path integral formalism,''
  Phys.\ Rev.\  {\bf D62}, 074501 (2000), 
  [hep-ph/9910306].
  % 
  % 
  \bibitem{Liu:2012ch} 
  K.~-F.~Liu, W.~-C.~Chang, H.~-Y.~Cheng and J.~-C.~Peng,
  %``Connected-Sea Partons,''
  Phys.\ Rev.\ Lett.\  {\bf 109}, 252002 (2012)
  [arXiv:1206.4339 [hep-ph]].  
  %
  \%
 \bibitem{Karsten:1980wd} 
  L.~H.~Karsten and J.~Smit,
  %``Lattice Fermions: Species Doubling, Chiral Invariance, and the Triangle Anomaly,''
  Nucl.\ Phys.\ B {\bf 183}, 103 (1981).
 %
 % 
  \bibitem{Lagae:1994bv} 
  J.~F.~Lagae and K.~F.~Liu,
  %``Finite mass corrections for sea quark matrix elements on the lattice,''
  Phys.\ Rev.\ D {\bf 52}, 4042 (1995)
  [hep-lat/9501007].  
 %  
%
\bibitem{Neuberger:1997fp} 
  H.~Neuberger,
  %``Exactly massless quarks on the lattice,''
  Phys.\ Lett.\ B {\bf 417}, 141 (1998)
  [hep-lat/9707022].
%
%
\bibitem{Luscher:1998pqa} 
  M.~Luscher,
  %``Exact chiral symmetry on the lattice and the Ginsparg-Wilson relation,''
  Phys.\ Lett.\ B {\bf 428}, 342 (1998)
  [hep-lat/9802011].
 %
 %
\bibitem{Hasenfratz:1998ri} 
  P.~Hasenfratz, V.~Laliena and F.~Niedermayer,
  %``The Index theorem in QCD with a finite cutoff,''
  Phys.\ Lett.\ B {\bf 427}, 125 (1998)
  [hep-lat/9801021].
%
%
\bibitem{Kikukawa:1998pd} 
  Y.~Kikukawa and A.~Yamada,
  %``Weak coupling expansion of massless QCD with a Ginsparg-Wilson 
  % fermion and axial U(1) anomaly,''
  Phys.\ Lett.\ B {\bf 448}, 265 (1999)
  [hep-lat/9806013];
%  \bibitem{Adams:1998eg} 
  D.~H.~Adams,
  %``Axial anomaly and topological charge in lattice gauge theory with overlap Dirac operator,''
  Annals Phys.\  {\bf 296}, 131 (2002)
  [hep-lat/9812003];
%  \bibitem{Fujikawa:1998if} 
  K.~Fujikawa,
  %``A Continuum limit of the chiral Jacobian in lattice gauge theory,''
  Nucl.\ Phys.\ B {\bf 546}, 480 (1999)
  [hep-th/9811235];
%  \bibitem{Suzuki:1998yz} 
  H.~Suzuki,
  %``Simple evaluation of chiral Jacobian with overlap Dirac operator,''
  Prog.\ Theor.\ Phys.\  {\bf 102}, 141 (1999)
  [hep-th/9812019].
%
%
\bibitem{Kikukawa:1998py} 
  Y.~Kikukawa and A.~Yamada,
  %``Axial vector current of exact chiral symmetry on the lattice,''
  Nucl.\ Phys.\ B {\bf 547}, 413 (1999)
  [hep-lat/9808026].
%
%\bibitem{Yang:2014sea} 
 % Y.~B.~Yang, Y.~Chen, A.~Alexandru, S.~J.~Dong, T.~Draper, M.~Gong, F.~X.~Lee and A.~Li {\it et al.},
  %``Charm and strange quark masses and $f_{D_s}$ from overlap fermions,''
 % arXiv:1410.3343 [hep-lat].
%
\bibitem{Gong:2013vja} 
  M.~Gong {\it et al.}  [XQCD Collaboration],
  %``Strangeness and charmness content of the nucleon from overlap 
  % fermions on 2+1-flavor domain-wall fermion configurations,''
  Phys.\ Rev.\ D {\bf 88}, no. 1, 014503 (2013)
  [arXiv:1304.1194 [hep-ph]].
  %
  %
  \bibitem{Franz:2000ee} 
  M.~Franz, M.~V.~Polyakov and K.~Goeke,
  %``Heavy quark mass expansion and intrinsic charm in light hadrons,''
  Phys.\ Rev.\ D {\bf 62}, 074024 (2000)
  [hep-ph/0002240].
%
%  
\bibitem{Liu:1995kb} 
  K.~F.~Liu,
  %``Comments on lattice calculations of proton spin components,''
  hep-lat/9510046; K.~F.~Liu, S.~J.~Dong, T.~Draper and W.~Wilcox,
  %``Pi N N and pseudoscalar form-factors from lattice QCD,''
  Phys.\ Rev.\ Lett.\  {\bf 74}, 2172 (1995)
  [hep-lat/9406007].
 %   
 %
\bibitem{Jaffe:1989jz} 
  R.~L.~Jaffe and A.~Manohar,
  %``The G(1) Problem: Fact and Fantasy on the Spin of the Proton,''
  Nucl.\ Phys.\ B {\bf 337}, 509 (1990).
%
%  
 \bibitem{Manohar:1990kr} 
  A.~V.~Manohar,
  %``Parton distributions from an operator viewpoint,''
  Phys.\ Rev.\ Lett.\  {\bf 65}, 2511 (1990) 
 %
 %
\bibitem{Chen:2008ag} 
  X.~-S.~Chen, X.~-F.~Lu, W.~-M.~Sun, F.~Wang and T.~Goldman,
  %``Spin and orbital angular momentum in gauge theories: Nucleon spin structure 
  % and multipole radiation revisited,''
  Phys.\ Rev.\ Lett.\  {\bf 100}, 232002 (2008)
  [arXiv:0806.3166 [hep-ph]].
%
%
\bibitem{Chen:2009mr} 
  X.~-S.~Chen, W.~-M.~Sun, X.~-F.~Lu, F.~Wang and T.~Goldman,
  %``Do gluons carry half of the nucleon momentum?,''
  Phys.\ Rev.\ Lett.\  {\bf 103}, 062001 (2009)
  [arXiv:0904.0321 [hep-ph]].
 %
 %
\bibitem{Wakamatsu:2010qj} 
  M.~Wakamatsu,
  %``On Gauge-Invariant Decomposition of Nucleon Spin,''
  Phys.\ Rev.\ D {\bf 81}, 114010 (2010)
  [arXiv:1004.0268 [hep-ph]].
%
%
\bibitem{Hatta:2011zs} 
  Y.~Hatta,
  %``Gluon polarization in the nucleon demystified,''
  Phys.\ Rev.\ D {\bf 84}, 041701 (2011)
  [arXiv:1101.5989 [hep-ph]].
 %
 %
 \bibitem{Cho:2010cw} 
  Y.~M.~Cho, M.~-L.~Ge and P.~Zhang,
  %``Nucleon Spin in QCD: Old Crisis and New Resolution,''
  Mod.\ Phys.\ Lett.\ A {\bf 27}, 1230032 (2012)
  [arXiv:1010.1080 [nucl-th]].
%
%
 \bibitem{Leader:2013jra} 
  E.~Leader and C.~Lorcé,
  %``The angular momentum controversy: What?s it all about and does it matter?,''
  Phys.\ Rept.\  {\bf 541}, 163 (2014)
  [arXiv:1309.4235 [hep-ph]].
  %
  %
\bibitem{cdg89}
C. Cohen-Tannoudji, J. Dupont-Roc, and G. Grynberg, \underline{Photons and Atoms}
(Wiley, New York  1989).
 %
 % 
 \bibitem{en94}
S.J. van Enk and G. Nienhuis, J. Mod. Opt. {\bf 41}, 963 (1994); 
S.J. van Enk and G. Nienhuis, Europhys. Lett. {\bf 25}, 497 (1994). 
%
%
\bibitem{Bliokh:2012zr} 
  K.~Y.~Bliokh, A.~Y.~Bekshaev and F.~Nori,
  %``Dual electromagnetism: Helicity, spin, momentum, and angular momentum,''
  New J.\ Phys.\  {\bf 15}, 033026 (2013)
  [arXiv:1208.4523 [physics.optics]];{\it ibid} 073022 (2013).
%
% 
\bibitem{Bliokh:2014ara} 
  K.~Y.~Bliokh, J.~Dressel and F.~Nori,
  %``Conservation of the spin and orbital angular momenta in electromagnetism,''
  New J.\ Phys.\  {\bf 16}, no. 9, 093037 (2014)
  [arXiv:1404.5486 [physics.optics]].
 %
 % 
\bibitem{beth36}
R.A. Beth, Phys. Rev. {\bf 50}, 115 (1936).
%
%
\bibitem{bsw92}
L. Allen, M.W. Beijersbergen, R.J.C. Spreeuw, and J.P. Woerdman, Phys. Rev. {\bf A 45},
8185 (1992); S.J. van Enk and G. Nienhuis, Opt. Commun. {\bf 94}, 147 (1992);
M.W. Beijersbergen, L. Allen, , H.E.L.O. van der Veen, and J.P. Woerdman, Opt. Commun. {\bf 96}, 
123 (1993).
%
%
\bibitem{Ji:2013fga} 
  X.~Ji, J.~H.~Zhang and Y.~Zhao,
  %``Physics of the Gluon-Helicity Contribution to Proton Spin,''
  Phys.\ Rev.\ Lett.\  {\bf 111}, 112002 (2013)
  [arXiv:1304.6708 [hep-ph]].  
%
%  
\bibitem{yl14}
Y.B. Yang and K.F. Liu, under preparation.
%
%
 \bibitem{Sufian:2014jma} 
  R.~S.~Sufian, M.~J.~Glatzmaier, Y.~B.~Yang, K.~F.~Liu and M.~Sun,
  %``Glue Spin $S_G$ in The Longitudinally Polarized Nucleon,''
  arXiv:1412.7168 [hep-lat]. 
%
%  
 \bibitem{Gamberg:2014zwa} 
  L.~Gamberg, Z.~B.~Kang, I.~Vitev and H.~Xing,
  %``Quasi-parton distribution functions: a study in the diquark spectator model,''
  Phys.\ Lett.\ B {\bf 743}, 112 (2015)
  [arXiv:1412.3401 [hep-ph]].  
 % 
 %
  
\end{thebibliography}
\end{document}